\documentclass[preprint,prl, superscriptaddress,longbibliography]{revtex4-2}

\usepackage{graphicx}
\usepackage{color}
\usepackage{amsfonts}

\usepackage{isomath}
\usepackage{amsmath}
\usepackage{amsthm}
\usepackage{amsfonts}
\usepackage{amssymb}
\usepackage{fdsymbol}
\usepackage{braket}
\usepackage{gensymb}

\let\oldAA\AA
\renewcommand{\AA}{\text{\normalfont\oldAA}}

\begin{document}

\title{Quasiparticle gap renormalization driven by internal and external screening in a WS$_2$ device}
\author{Chakradhar Sahoo}
\email{sahoo1@phys.au.dk}
\affiliation{Department of Physics and Astronomy, Interdisciplinary Nanoscience Center, Aarhus University, 8000 Aarhus C, Denmark}
\author{Yann in $`$t Veld}
\affiliation{Institute for Molecules and Materials, Radboud University, 6525 AJ Nijmegen, the Netherlands}
\author{Alfred J. H. Jones}
\author{Zhihao Jiang}
\affiliation{Department of Physics and Astronomy, Interdisciplinary Nanoscience Center, Aarhus University, 8000 Aarhus C, Denmark}
\author{Greta Lupi}
\affiliation{Department of Applied Physics, Aalto University, 02150 Espoo, Finland}
\author{Paulina E.  Majchrzak}
\author{Kimberly Hsieh}
\affiliation{Department of Physics and Astronomy, Interdisciplinary Nanoscience Center, Aarhus University, 8000 Aarhus C, Denmark}
\author{Kenji~Watanabe}
\affiliation{Research Center for Electronic and Optical Materials, National Institute for Materials Science, 1-1 Namiki, Tsukuba 305-0044, Japan}
\author{Takashi~Taniguchi}
\affiliation{Research Center for Materials Nanoarchitectonics, National Institute for Materials Science,  1-1 Namiki, Tsukuba 305-0044, Japan}
\author{Philip Hofmann}
\affiliation{Department of Physics and Astronomy, Interdisciplinary Nanoscience Center, Aarhus University, 8000 Aarhus C, Denmark}
\author{Jill A.  Miwa}
\affiliation{Department of Physics and Astronomy, Interdisciplinary Nanoscience Center, Aarhus University, 8000 Aarhus C, Denmark}
\author{Yong P.  Chen}
\affiliation{Department of Physics and Astronomy, Interdisciplinary Nanoscience Center, Aarhus University, 8000 Aarhus C, Denmark}
\affiliation{Department of Physics and Astronomy and School of Electrical and Computer Engineering and Purdue Quantum Science and Engineering Institute, Purdue University, West Lafayette, IN 47907, USA}
\author{Malte Rösner}
\affiliation{Institute for Molecules and Materials, Radboud University, 6525 AJ Nijmegen, the Netherlands}
\author{Søren Ulstrup}
\email{ulstrup@phys.au.dk}
\affiliation{Department of Physics and Astronomy, Interdisciplinary Nanoscience Center, Aarhus University, 8000 Aarhus C, Denmark}

\begin{abstract}
The electronic band gap of a two-dimensional semiconductor within a device architecture is sensitive to variations in screening properties of adjacent materials in the device and to gate-controlled doping. Here,  we employ micro-focused angle resolved photoemission spectroscopy to separate band gap renormalization effects stemming from environmental screening and electron-doping during \textit{in situ} gating of a single-layer WS$_{2}$ device. The WS$_{2}$ is supported on hBN and contains a section that is exposed to vacuum and another section that is encapsulated by a graphene contact. We directly observe the doping-induced semiconductor-metal transition and band gap renormalization in the two sections of WS$_2$.  Surprisingly,  a larger band gap renormalization is observed in the vacuum-exposed section than in the graphene-encapsulated — and thus ostensibly better screened — section of the WS$_2$.  Using $GW$ calculations,  we determine that intrinsic screening due to stronger doping in vacuum exposed WS$_2$ exceeds the external environmental screening in graphene-encapsulated WS$_2$. 
\end{abstract}

\maketitle

Quasiparticle excitations and electronic band structures of two-dimensional (2D) semiconducting transition metal dichalcogenides (TMDCs) are widely tuneable via intrinsic and extrinsic perturbations to the Coulomb interaction \cite{Qiu:2013,WangG:2018}, which underpin the operation of 2D field effect transistors, light emitting diodes and quantum many-body simulators based on TMDCs \cite{britnell2012field,withers2015light,novoselov20162d,liu2021promises,Tang:2020,Zuocheng:2022}. The architechture of such devices employs hBN as dielectric and graphene as contacts that are either partially or fully encapsulating the TMDC in question, affecting the external dielectric screening and resulting band alignments, doping and band gap renormalization \cite{Ugeda2014,miwaelectronic2015,raja2017coulomb,Ulstrup:align2019,waldecker2019rigid,utama2019dielectric}. The possibility of tuning the charge carrier doping in the TMDC via electrostatic gating, optical excitation or alkali adsorption adds an additional internal contribution to the screening that significantly changes the electronic and optical properties \cite{zhangdirect2014,Chernikov:2015,Antonija-Grubisic-Cabo:2015aa,Zhang2016,Ulstrup:2016,katoch2018giant,Madeo:2020,Reutzel31122024,ulstrup2024observation,NHofmann:2025}. To our knowledge, the detailed interplay of external and internal screening channels has so far only been discussed with respect to superconducting instabilities~\cite{SCScrPRB2016} as well as in charge-neutral photodoped TMDCs~\cite{steinhoff_exciton_2017}.

\begin{figure*}[t!]
\begin{center}
\includegraphics[scale=1.0]{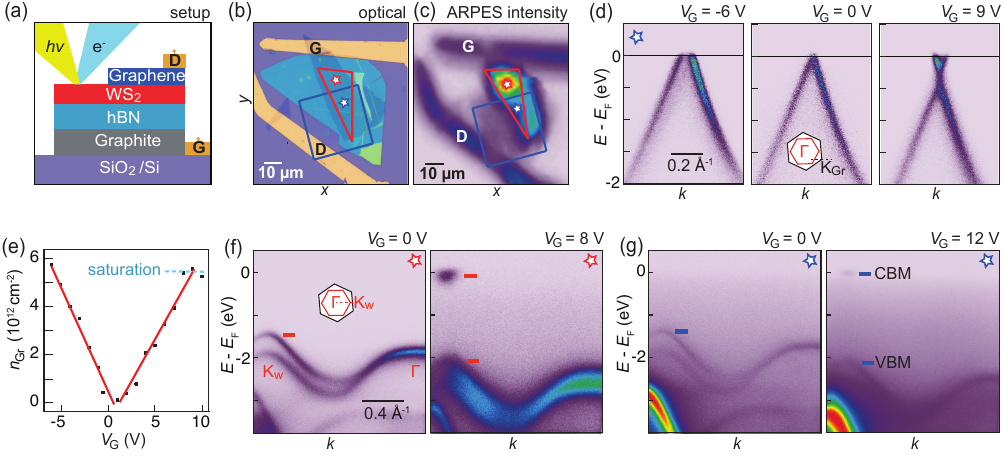}
\caption{(a) Simplified schematic of microARPES experiment on a gated SL WS$_2$ device.  Gate and drain electrodes are labelled as G and D.  (b) Optical image of the device.  Note that gate and drain electrodes have been pre-patterned and are therefore located under graphite and graphene, respectively.  The setup is equivalent to the simplified diagram in (a).  (c) Energy- and momentum-integrated map of ARPES intensity from the device region.  The red triangular outline and the blue square outline in (b)-(c) demarcate the SL WS$_2$ and graphene flakes,  respectively.  (d) ARPES spectra of the graphene Dirac cone at the given gate voltages ($V_{G}$) obtained along the cut through $\mathrm{K}_{\mathrm{Gr}}$ marked on the BZs of graphene (black hexagon) and SL WS$_2$ (red hexagon) in the inset.  The measurement was performed in the spot marked by a blue star in (c). (e) Graphene carrier density $n_{\mathrm{Gr}}$ as a function of $V_{\mathrm{G}}$ obtained from ARPES spectra of the Dirac cone as shown in (d) \cite{SMAT}. Red full lines represent linear fits. The saturation level for electron-doping of the graphene is demarcated by a blue dashed line.  (f)-(g) ARPES spectra of SL WS$_2$ at the stated values of $V_{G}$ measured along $\mathrm{\Gamma}-\mathrm{K}_{\mathrm{W}}$.  The ARPES cut direction is shown on the BZs in the inset.  The energies of the VBM and CBM obtained from EDC fits are marked by red and blue ticks for spectra obtained in the two regions of the device demarcated by (f) red and (g) blue stars in (b)-(c) \cite{SMAT}.}
\label{Fig:1}
\end{center}
\end{figure*}

We disentangle external and internal screening channels using a single-layer (SL) WS$_2$ device, where the SL WS$_2$ is supported on hBN. It is partially exposed to vacuum and partially encapsulated by graphene, as sketched in Fig.  \ref{Fig:1}(a). This architecture provides access to two distinct environmental screening scenarios within the same SL WS$_2$ flake while we are simultaneously able to tune the carrier concentration, and thus the internal screening, using electrostatic gating by applying a gate voltage ($V_{\mathrm{G}}$) to a bottom graphite electrode. 

The impact of screening is determined by directly observing the quasiparticle band structure during \textit{in situ} gating using angle-resolved photoemission spectroscopy with micrometer spatial resolution (microARPES) \cite{nguyen2019visualizing,hofmann2021accessing,nguyen2021field,graham2024conduction, graham2024band}. The TMDC device is prepared via mechanical exfoliation and dry transfer methods \cite{caldwell2010technique,yi2015review,SMAT}. The device is presented in the optical micrograph in Fig.  \ref{Fig:1}(b) where red and blue outlines demarcate SL WS$_2$ and graphene flakes,  respectively. The microARPES measurements were performed at the SGM4 beamline of the ASTRID2 light source at Aarhus University,  Denmark \cite{Jones:2025}, at room temperature, and at a photon energy of 55 eV.  The energy and momentum resolution were set to 35~meV and 0.01 $\AA^{-1}$,  respectively.  A capillary mirror was used to focus the beam to a spot size of 4 $\mu$m.  By raster scanning the beam over the area seen in Fig.  \ref{Fig:1}(b) we collect an ($x$,$y$)-dependent map of the ($E$,$k$)-resolved ARPES intensity,  as shown in Fig.  \ref{Fig:1}(c),  enabling us to single out vacuum exposed and graphene-encapsulated SL WS$_2$ areas for $V_{\mathrm{G}}$-dependent measurements.  In the following,  we focus our measurements on two such areas, which have been demarcated by red and blue stars in Figs.  \ref{Fig:1}(b)-(c).

In order to understand the $V_{\mathrm{G}}$-induced charge density of the functional device, and to extract the Fermi level ($E_{\mathrm{F}}$) reference,  we start by measuring ($E$,$k$)-dependent ARPES spectra of the graphene Dirac cone.  Spectra obtained at $V_{\mathrm{G}}$ values of -6,  0 and 9 V are presented in Fig.  \ref{Fig:1}(d).  These spectra demonstrate the filling of the Dirac cone as $V_{\mathrm{G}}$ is increased.   By extracting the graphene carrier density $n_{\mathrm{Gr}}$ at each measured $V_{\mathrm{G}}$ \cite{SMAT},  we find that saturation for electron-doping occurs at 8 V and sets an upper limit for $n_{\mathrm{Gr}}$ at around $5\times 10^{12}$ cm$^{-2}$,  as shown in Fig.  \ref{Fig:1}(e).  As we will show later,  we find that the saturation of electron-doping in graphene occurs simultaneously with the filling of the vacuum exposed SL WS$_2$ conduction band minimum (CBM) at the point where the SL WS$_2$ has become fully metallic and adheres to the Fermi level in graphene.

\begin{figure} [t] 
\begin{center}
\includegraphics[width=0.49\textwidth]{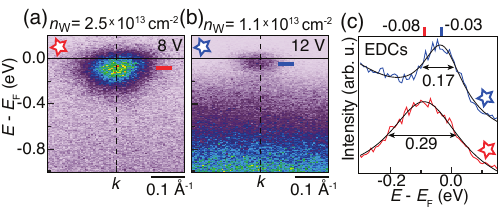}
\caption{(a)-(b) Close-up of ARPES spectra of the CBM region in (a) vacuum exposed and (b) graphene-encapsulated SL WS$_2$ with estimates of electron density $n_{\mathrm{W}}$.  (c) EDCs extracted along the dashed vertical line at $\mathrm{K}_{\mathrm{W}}$ in (a)-(b).  Fits to a function composed of a single Lorentzian peak on a linear background multiplied by the Fermi-Dirac distribution are overlaid.  The blue and red ticks demarcate the fitted peak positions.  Double-headed arrows demarcate the full width at half maximum of the fitted Lorentzian peaks.  Values are stated in units of eV and have error bars of $\pm 0.02$ eV.}
\label{Fig:2}
\end{center}
\end{figure}

Turning to the vacuum exposed section of SL WS$_2$ in the device, we present the $E(k)$ band dispersion along $\mathrm{\Gamma}-\mathrm{K}_{\mathrm{W}}$ in Fig.  \ref{Fig:1}(f).  Note that the Fermi level is always referenced to the Fermi energy of the graphene top contact.  We present spectra at $V_{\mathrm{G}} = 0$~V and at the electron-doping saturation level of $V_{\mathrm{G}} = 8$~V determined on graphene.  At $V_{\mathrm{G}} = 0$~V,  we observe the valence band maximum (VBM) to be located at $\mathrm{K}_{\mathrm{W}}$ at an energy of -1.48~eV and with a spin-orbit splitting of 0.42~eV,  consistent with the expected band structure of SL WS$_2$ \cite{Manzeli:2017}.  At $V_{\mathrm{G}} = 8$~V, the VBM is shifted by 0.61~eV to lower energies,  the CBM has shifted below $E_{\mathrm{F}}$ and a substantial energy broadening of the bands has occurred.  We find that the linewidth of energy distribution curves (EDCs) through the VBM nearly doubles compared to the situation at $V_{\mathrm{G}} = 0$~V \cite{SMAT}.  Such broadening primarily results from photo-induced currents in the SL WS$_2$ that leads to electrostatic potential variations across the 4~$\mu$m length scale of the light spot,  which causes  inhomogeneous doping and energy-shifting of the bands \cite{nguyen2019visualizing,graham2024conduction, graham2024band}.

We do not encounter this problem in the graphene-encapsulated section of the SL WS$_2$ device, as we do not observe any energy broadening under the graphene \cite{SMAT}.  In this case,  the photoemitted charge in SL WS$_2$ is efficiently replenished via the highly conductive graphene contact,  which greatly reduces inhomogeneous potential variations across the light spot.  ARPES spectra are presented in Fig.  \ref{Fig:1}(g).  The reduced intensity and higher background compared with the spectra of bare SL WS$_2$ in Fig.  \ref{Fig:1}(f) are caused by inelastic photoelectron scattering in the overlying graphene.  At $V_{\mathrm{G}} = 0$~V,  the VBM is located at -1.40~eV and thus shifted further towards $E_{\mathrm{F}}$ than in the vacuum exposed SL WS$_2$ section of the device. This shift is due to the enforced band alignment upon interfacing SL WS$_2$ with graphene.  Tuning to $V_{\mathrm{G}} = 12$~V leads to a downwards VBM shift of 0.80 eV and a faint CBM state at $E_{\mathrm{F}}$.  We do not observe additional energy and momentum broadening of the SL WS$_2$ bands in this part of the device when we increase $V_{\mathrm{G}}$ to the maximum possible value of 12 V before gate leakage through hBN sets in \cite{SMAT}.

Detailed ARPES spectra of the SL WS$_2$ CBM region are presented in Fig.  \ref{Fig:2} for the two sections of the device.  Fits of EDCs through K$_\mathrm{W}$ reveal CBM energies of -0.08~eV and -0.03~eV in the vacuum exposed and graphene-encapsulated parts, respectively (see Fig.  \ref{Fig:2}(c)) \cite{SMAT,Nechaev:2009,Shiang:2015}.  The EDC linewidth is nearly doubled in bare SL WS$_2$ due to extrinsic broadening from the photocurrent.  Considering a quadratic dispersion,  we estimate the SL WS$_2$ carrier concentration $n_{\mathrm{W}}$ from the fitted CBM. The values of $n_{\mathrm{W}}$ are stated in Figs.  \ref{Fig:2}(a)-(b) and reveal a doping that is more than twice as large in vacuum exposed SL WS$_2$ at $V_{\mathrm{G}} = 8$~V than in the graphene-encapsulated part at $V_{\mathrm{G}} = 12$~V.  

\begin{figure}[t!]
\begin{center}
\includegraphics[width=0.49\textwidth]{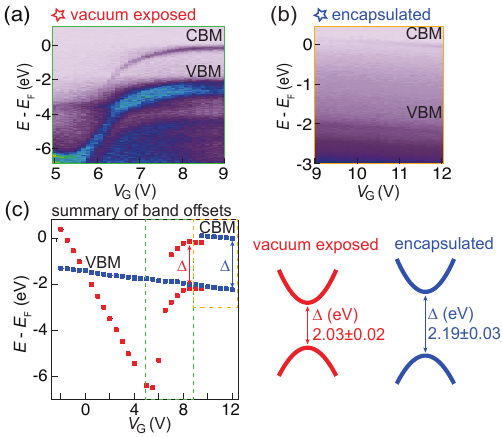}
\caption{(a)-(b) $V_{\mathrm{G}}$-dependent sweeps of EDC at $\mathrm{K}_{\mathrm{W}}$ of SL WS$_{2}$ measured on (a) vacuum exposed and (b) graphene-encapsulated sections. The $V_{\mathrm{G}}$ ranges are restricted to where intensity from the CBM emerges in the two sections.  (c) Mean energies of VBM and CBM positions of SL WS$_{2}$ for vacuum exposed (red markers) and graphene-encapsulated (blue markers) sections across a wide $V_{\mathrm{G}}$ range between -2~V and 12~V.  Note that measurements were performed down to $V_{\mathrm{G}} = -6$~V,  which are not shown since the bands merely exhibit a monotonic shift with gate voltage in this range.  Dashed green and orange boxes demarcate the $V_{\mathrm{G}}$- and energy-ranges presented in (a) and (b). Double-headed arrows demarcate band gaps $\mathrm{\Delta}$. Mean values and standard deviation of the gaps in the two sections of SL WS$_2$ are stated in the band diagrams.}
\label{Fig:3}
\end{center}
\end{figure}

We determine the $V_{\mathrm{G}}$-dependence of VBM and CBM positions in order to track the direct band gap in the two sections of the device. On vacuum exposed SL WS$_2$, we observe a rigid linear shift to lower energies of the entire photoemission spectrum of WS$_2$ that tracks the back-gate potential in the range from 0 to 5~V (see \cite{SMAT}). Above $V_{\mathrm{G}} = 5$~V, the conductivity of the vacuum exposed section of the WS$_2$ starts to increase, which we track in Fig. \ref{Fig:3}(a) via a $V_{\mathrm{G}}$-dependent sweep of an EDC at $\mathrm{K}_{\mathrm{W}}$. The increasing conductivity with $V_{\mathrm{G}}$ causes the WS$_2$ Fermi energy to shift back towards the graphene reference level. Above $V_{\mathrm{G}} = 8$~V, no further shifting occurs and we observe the CBM straddling the Fermi level referenced to the graphene, consistent with a stable conducting WS$_2$ layer. The complete semiconductor-to-metal transition of the vacuum exposed WS$_2$ at $V_{\mathrm{G}} = 8$~V occurs concomitantly with the saturation doping of graphene as seen in Fig. \ref{Fig:1}(e). 

In graphene-encapsulated WS$_2$ we observe a monotonous downwards shift of the VBM of 0.58 eV from 0 to 9 V (see \cite{SMAT}), consistent with a continuous doping effect without shifting of the WS$_2$ Fermi energy with respect to the graphene reference level. A $V_{\mathrm{G}}$ sweep from 9 to 12 V of an EDC at $\mathrm{K}_{\mathrm{W}}$ in this section of the device is shown in Fig. \ref{Fig:3}(b), highlighting the onset of CBM population at 9.5 V and a continuous downwards shift as $V_{\mathrm{G}}$ further increases. 

By repeatedly ramping $V_{\mathrm{G}}$ up and down between negative and positive voltage values, we continuously and reversibly tune the band positions, which we extract from EDCs as in Figs. \ref{Fig:3}(a)-(b) in both sections of the device, as summarized in Fig. \ref{Fig:3}(c).  Note that increasing energy broadening makes extraction of these band positions impossible above 10 V on vacuum exposed WS$_2$. From the extracted VBM and CBM positions we obtain estimates of the direct band gap $\mathrm{\Delta}$ in the two sections of the device. Surprisingly, we observe that $\mathrm{\Delta} = (2.03 \pm 0.02)$~eV in vacuum exposed WS$_2$ while we find that $\mathrm{\Delta} = (2.19 \pm 0.03)$~eV in graphene-encapsulated WS$_2$ across the measured $V_{\mathrm{G}}$ range (see band diagrams in Fig. \ref{Fig:3}(c)). Subtle changes of $\mathrm{\Delta}$ with $V_{\mathrm{G}}$ occur, but they are within the experimental uncertainty \cite{SMAT}. Considering only the additional environmental screening from the graphene layer one would expect the opposite trend of what we observe: The overall Coulomb interaction should be suppressed in the graphene-encapsulated area yielding a reduced local quasiparticle gap \cite{Ugeda2014,raja2017coulomb,waldecker2019rigid,steinke2020coulomb}. 

In a more detailed analysis, the vacuum exposed section of the device should be understood as a situation with a small environmental screening but large electron doping level, while the graphene-encapsulated section corresponds to a large environmental screening and small electron doping level. To disentangle these internal and external screening channels at finite charge doping by theoretical means, we performed ab initio model $G_0W_0$ calculations for WS$_{2}$ under the influence of electron doping ($n_{\mathrm{W}} > 0$) and environmental dielectric screening ($\varepsilon_{\mathrm{env}} > 1)$ \cite{SMAT,rosner2015wannier,aryasetiawan2004frequency-dependent,kresse1996efficient,kresse1999ultrasoft,blochl1994projector,perdew1996generalized,haastrup2018computational,gjerding2021recent,florian2018dielectric,parcollet2015triqs,strand2023tprf,aryasetiawan1996multiple,kas2014cumulant,caruso2015band}.

\begin{figure} [t] 
\begin{center}
\includegraphics[width=0.49\textwidth]{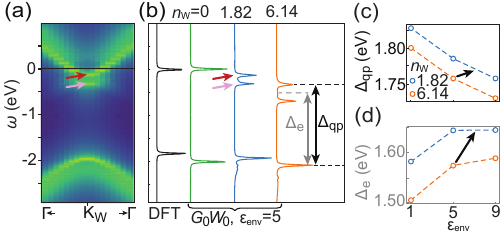}
\caption{
(a) Interacting $G_0W_0$ spectral function of SL WS$_{2}$ at low electron doping ($n_{\mathrm{W}} = 1.82 \times 10^{13}$ cm$^{-2}$) and with a dielectric environmental screening of $\varepsilon_{\mathrm{env}} = 5$. (b) EDCs at $\mathrm{K}_{\mathrm{W}}$ obtained from DFT (black) and within $G_0W_0$ (green). The $G_0W_0$ EDCs are shown for $\varepsilon_{\mathrm{env}} = 5$ and at the stated doping levels $n_{\mathrm{W}}$ in units of $10^{13}$ cm$^{-2}$. Red and magenta arrows indicate quasiparticle (red) and plasmon satellite (magenta) peak positions. In the undoped cases we set the CBM to $\omega = 0$~eV. (c)-(d) Quasiparticle gap $\mathrm{\Delta}_{\mathrm{qp}}$ and the effective gap $\mathrm{\Delta}_{\mathrm{e}}$ demarcated by double-headed arrows in (b) as a function of $\varepsilon_{\mathrm{env}}$ for the stated doping levels. The arrows in (c) and (d) indicate situations in which increasing the environmental screening can increase corresponding gaps. Spin-orbit coupling effects are neglected in all cases. 
}
\label{Fig:4}
\end{center}
\end{figure}

In Fig. \ref{Fig:4}(a) we show an example of a $G_0W_0$ spectral function for a small electron doping ($n_{\mathrm{W}} \approx 1.82 \times 10^{13}$ cm$^{-2}$) and an intermediate environmental screening ($\varepsilon_{\mathrm{env}} = 5$) around $\mathrm{K}_{\mathrm{W}}$. The quasiparticle VBM $E^{(\mathrm{v})}_{\mathrm{qp}}(\mathrm{K}_{\mathrm{W}})$ occurs around $-2$~eV (spin-orbit coupling effects are neglected) and the CBM is occupied, such that $E^{(\mathrm{c})}_{\mathrm{qp}}(\mathrm{K}_{\mathrm{W}})$ (red arrow in Fig. \ref{Fig:4}(a)) is below 0~eV. As a result of the finite doping, low-energetic (``acoustic'') plasmons are formed, which can dress the quasiparticles yielding plasmon satellite features shifted by a plasma energy $\omega_{\mathrm{pl}}(\mathrm{K}_{\mathrm{W}})$ (see purple arrow in Fig. \ref{Fig:4}(a)) \cite{steinke2020coulomb,Vito:2024,CarusoPRB}. In Fig. \ref{Fig:4}(b) we show the corresponding EDC taken at $\mathrm{K}_{\mathrm{W}}$ next to EDCs for different doping levels and a standard undoped DFT calculation for reference. In the following, we differentiate between the quasiparticle gap $\mathrm{\Delta}_{\mathrm{qp}} = E^{(\mathrm{c})}_{\mathrm{qp}}(\mathrm{K}_{\mathrm{W}}) - E^{(\mathrm{v})}_{\mathrm{qp}}(\mathrm{K}_{\mathrm{W}})$, and an ``effective gap'' that we define as $\mathrm{\Delta}_{\mathrm{e}} = E^{(\mathrm{c})}_{\mathrm{qp}}(\mathrm{K}_{\mathrm{W}}) - \frac{1}{2}\omega_{\mathrm{pl}}(\mathrm{K}_{\mathrm{W}}) - E^{(\mathrm{v})}_{\mathrm{qp}}(\mathrm{K}_{\mathrm{W}})$, which are indicated by double-headed arrows in Fig. \ref{Fig:4}(b). In Figs. \ref{Fig:4}(c)-(d) we further show how $\mathrm{\Delta}_{\mathrm{qp}}$ and $\mathrm{\Delta}_{\mathrm{e}}$ are affected by $n_{\mathrm{W}}$ and $\varepsilon_{\mathrm{env}}$.

From Fig. \ref{Fig:4}(c) we see that $\mathrm{\Delta}_{\mathrm{qp}}$ decreases by only about 25~meV with $n_{\mathrm{W}}$ increasing from $1.82 \times 10^{13}$ cm$^{-2}$ to $6.14 \times 10^{13}$ cm$^{-2}$. This results from just minor changes to the density of states at the Fermi level, such that the Thomas-Fermi screening length and the final Coulomb interaction only mildly change. Increasing $\varepsilon_{\mathrm{env}}$ from $1$ to $9$ decreases $\mathrm{\Delta}_{\mathrm{qp}}$ by more than 60~meV. This trend is mainly controlled by the (effective) thickness of WS$_2$ and thus by its distance to the dielectric environment. It is the ratio of the Thomas-Fermi screening length to this (effective) thickness that mostly controls the relative impact of $n_{\mathrm{W}}$ and $\varepsilon_{\mathrm{env}}$ to $\mathrm{\Delta}_{\mathrm{qp}}$~\cite{SCScrPRB2016}: The smaller the density of states at the Fermi level (the smaller the Thomas-Fermi screening length) and the smaller the effective thickness of the 2D material, the larger the impact of $\varepsilon_{\mathrm{env}}$ to $\mathrm{\Delta}_{\mathrm{qp}}$.

Combining the contributions of electron doping and environmental screening to $\mathrm{\Delta}_{\mathrm{qp}}$, a situation can be realized where $\mathrm{\Delta}_{\mathrm{qp}}$ increases while $\varepsilon_{\mathrm{env}}$ increases if $n_{\mathrm{W}}$ decreases -- as indicated by the arrow in Fig. \ref{Fig:4}(c). This is the situation we find in our experiment when going from vacuum exposed WS$_2$ (small $\varepsilon_{\mathrm{env}}$, large $n_{\mathrm{W}}$) to graphene-encapsulated WS$_2$ (large $\varepsilon_{\mathrm{env}}$, small $n_{\mathrm{W}}$) and we can thus explain the surprising observation of a larger $\mathrm{\Delta}_{\mathrm{qp}}$ in encapsulated WS$_2$. There is a discrepancy between theory and experiment as the maximum gap change in the theory is merely around 25~meV while the experiment reveals an effect larger than 100~meV. This might be explained by the approximations in the many-body model \cite{SMAT}. It might also be explained if we take the non-quasiparticle $G_0W_0$ plasmon satellite into account and analyze $\mathrm{\Delta}_{\mathrm{e}}$ instead of $\mathrm{\Delta}_{\mathrm{qp}}$. In this case, we find situations with much larger increasing $\mathrm{\Delta}_{\mathrm{e}}$ upon increasing $\varepsilon_{\mathrm{env}}$ (see arrow in Fig. \ref{Fig:4}(d)), which is in line with our experimental result. We note that these plasmon satellites have been observed before in electron doped TMDCs~\cite{ulstrup2024observation,CarusoPRB}, but are not resolved here due to a combination of smaller doping levels and thermal broadening, such that we measure the center energy between the individual quasiparticle peak and plasmon satellite features in our ARPES EDCs.

Our minimal many-body modeling reveals an intriguing interplay between external and internal screening channels (here enabled by the relatively low doping level), that is responsible for larger quasiparticle gaps in areas with large external screening but reduced doping levels and smaller quasiparticle gaps in areas with smaller external screening but larger doping levels. The ARPES observations are thus driven by modulated screening effects uncovering the relevance of spatially varying many-body effects in device architectures based on layered materials. \cite{Zenodo}

\section{Acknowledgements}
C.S.  acknowledges Marie Sklodowska-Curie Postdoctoral Fellowship (project 101059528 MaPWave).  The work was funded/co-funded by the European Union (ERC grant EXCITE with project number 101124619). Views and opinions expressed are
however those of the author(s) only and do not necessarily reflect those of the European Union or the European
Research Council.  Neither the European Union nor the granting authority can be held responsible for them. The authors acknowledge funding from the Novo Nordisk Foundation (Project Grant NNF22OC0079960, and NFF23OC0085585),  VILLUM FONDEN under the Villum Investigator Program (Grant. No. 25931), the Danish Council for Independent Research (Grant Nos.  DFF-9064-00057B,  DFF-6108-00409, DFF-4258-00002B and 1026-00089B), and the Aarhus University Research Foundation.  Y.V. and M.R. acknowledge support from the Dutch Research Council (NWO) via the “TOPCORE" consortium. K.W. and T.T. acknowledge support from the JSPS KAKENHI (Grant Numbers 21H05233 and 23H02052) , the CREST (JPMJCR24A5), JST and World Premier International Research Center Initiative (WPI), MEXT, Japan. C.S.  acknowledges Tony F.  Heinz for helpful discussions.

\newpage

\section{Supporting Information}

\section{I.  Device Preparation}

Flakes of SL WS$_{2}$ were exfoliated and integrated in a functional device using the following fabrication steps:

\begin{itemize}

\item[(a)] A bulk WS$_2$ crystal with dimensions on the order of a millimetre was picked up with a piece of scotch tap and stamped on blue tape (nitto tape).  In parallel,  a polydimethylsiloxane (PDMS) Gel-Pak (PF-40x40-0170-X4) of ($10 \times 50$) mm$^2$ was placed on a glass slide and cut into pieces of ($10 \times 2$) mm$^2$. The prepared blue tape was used to exfoliate WS$_2$ on the PDMS pieces, which were subsequently inspected with an optical microscope to identify single layers.

\item[(b)] A Gel-Pak piece with a large SL WS$_2$ flake was singled out to transfer the SL WS$_2$ on Si/SiO$_2$ (see I in Fig.  \ref{S1}).  Prior to the transfer,  the Si/SiO$_2$ substrate was cleaned in an oxygen plasma for 6 minutes.  During the transfer, the selected Gel-Pak was fixed on the corner of a glass slide and then mounted upside down on a micromanipulator (see II in Fig.  \ref{S1}). The Gel-Pak was then gently touched with the Si/SiO$_2$ at 30 $^{\circ}$C,  which was then heated to 60 $^{\circ}$C, and finally cooled back down to 30 $^{\circ}$C, allowing separation from the Gel-Pak and SL WS$_{2}$ transferred on Si/SiO$_2$ (see III in Fig.  \ref{S1}).  A polycarbonate (PC) film on a PDMS stamp was prepared on a glass slide and mounted in the micromanipulator (see 1 in Fig.  \ref{S1}).  The PC and WS$_{2}$ flake were brought into contact at 70 $^{\circ}$C, followed by heating to 90 $^{\circ}$C, and then cooling to 70 $^{\circ}$C, which results in the SL WS$_{2}$ being transferred to the PC/PDMS stamp (see 2 in Fig.  \ref{S1}). 

\item[(c)] Flakes of hBN were exfoliated directly on Si/SiO$_2$ at room temperature using scotch tape.  A suitable flake with a thickness of 14 nm was identified using an optical microscope. The flake was then picked up with the stamp using the procedure in (b) at a temperature of 120 $^{\circ}$C (see 3 in Fig.  \ref{S1}). 

\item[(d)] Graphite was exfoliated and a flake with a thickness around 20 nm was picked up with the stamp following the same procedure as for hBN in (c) (see 4 in Fig.  \ref{S1}).

\begin{figure*} [t!]
	\begin{center}
		\includegraphics[width=0.9\textwidth]{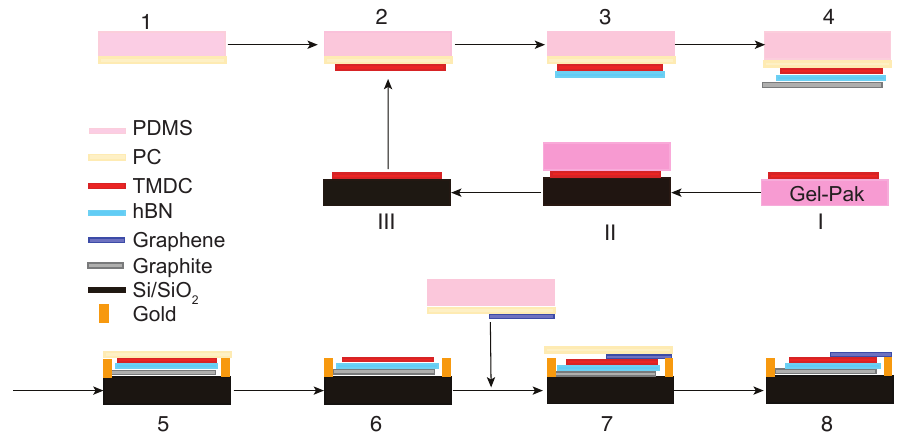}
		\caption{Processing steps for making the gated SL WS$_2$ device (see text for details). }
		\label{S1}
	\end{center}
\end{figure*}

\item[(e)] The stack was transferred to a Si/SiO$_2$ substrate with 50 nm thick pre-patterned gold electrodes,  which was prepared by photolithography.  The transfer was carried out at a temperature of 180 $^{\circ}$C,  causing the PC to melt and the stack to release from the PDMS stamp (see 5 in Fig.  \ref{S1}). 

\item[(f)] The stack was rinsed with chloroform,  acetone and isopropanol to remove PC and residues, which was followed by annealing at 300 $^{\circ}$C for 3 hours in the presence of $5\%$ H$_{2}/$Ar gas (see 6 in Fig.  \ref{S1}). 

\item[(g)]  Graphene was exfoliated on Si/SiO$_2$ and picked up with another PC/PDMS stamp following similar steps as for hBN and graphite. The graphene was then aligned and transferred on the device using the same procedure as described in (e) (see 7 in Fig.  \ref{S1}).  Our microARPES measurements revealed a twist angle of 25$^{\circ}$ between graphene and SL WS$_{2}$.

\item[(h)] The final stack was rinsed and annealed following the procedure in step (f) again (see 8 in Fig.  \ref{S1}).

\end{itemize}

The cleaned stack was mounted on a chip carrier (CSB00815) and wire-bonded.  The chip carrier was plugged into a custom-made sample plate that allowed \textit{in situ} electrical control in the ultra-high vacuum microARPES end station.  The gate voltage was set using a Keithley 2450 source meter.  Prior to measurements, the device was annealed over night at 200 $^{\circ}$C in the end station.  

\newpage

\section{II.  Extraction of Graphene Carrier Density}

\begin{figure*} [b!]
	\begin{center}
		\includegraphics[width=1\textwidth]{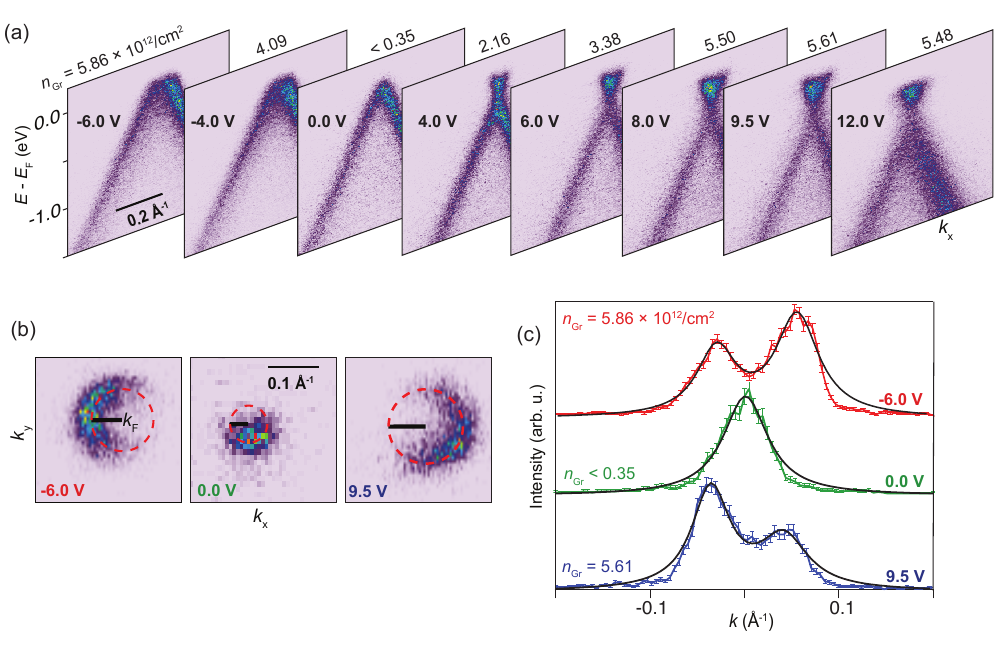}
		\caption{(a).  ARPES spectra of graphene Dirac cone as a function of $V_{\mathrm{G}}$.  The estimated carrier density is stated for each spectrum.  (b) Constant energy surfaces at $E_{\mathrm{F}}$ for the stated gate voltages.  (c) MDCs (colored curves) extracted across the Fermi surfaces in (b) with fits (black curves) to two Lorentzian peaks on a linear background. }
		\label{S2}
	\end{center}
\end{figure*}

The carrier density $n_{\mathrm{Gr}}$ in the graphene top contact is extracted as a function of $V_{\mathrm{G}}$ by fitting the Fermi wavevector $k_{\mathrm{F}}$ in $V_{\mathrm{G}}$-dependent ARPES spectra of the Dirac cone as shown in Fig.  \ref{S2}(a).  For a subset of $V_{\mathrm{G}}$ values we measure the circular Fermi surface and fit a circular contour with radius $k_{\mathrm{F}}$ as seen in Fig.  \ref{S2}(b).  In addition,  we extract momentum distribution curves (MDCs) at $E_{\mathrm{F}}$ and fit a function composed of two Lorentzian peaks on a linear background,  as shown in Fig.  \ref{S2}(c).  The distance between the fitted peaks corresponds to $2k_{\mathrm{F}}$.  In the low doping regime,  around $V_{\mathrm{G}} = 0$ V,  it is not possible to resolve two separate peaks at $E_{\mathrm{F}}$,  as seen via the green curve in Fig.  \ref{S2}(c).  We instead fit MDC cuts of the dispersion at lower energies and linearly extrapolate the branches to $E_{\mathrm{F}}$ where we then obtain $2k_{\mathrm{F}}$.  Combining these methods we obtain $n_{\mathrm{Gr}} = k_{\mathrm{F}}^2/\pi$ at each value of $V_{\mathrm{G}}$ directly from the ARPES dispersion of graphene.  We find that the lowest value of carrier density we can reliably estimate is 0.35$\times$10$^{12}$ cm$^{-2}$. Note that the extrapolation method contributes to the increased scatter of the points with respect to the linear fits seen for small $V_{\mathrm{G}}$-values in Fig.  1(e) of the main text.

\newpage

\section{III.  Analysis of Valence band positions and linewidths}

\begin{figure*} [h!]
	\begin{center}
		\includegraphics[width=0.5\textwidth]{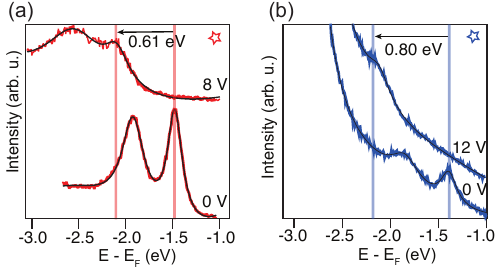}
		\caption{(a)-(b) EDC analysis of valence band at K$_{\mathrm{W}}$ for (a) vacuum exposed and (b) graphene-encapsulated SL WS$_2$.  The EDCs (colored curves) are fitted to a function consisting of Lorentzian peaks on a polynomial background.  The VBM determined by the fit is demarcated by vertical bars.  Shifts of the VBM from $V_{\mathrm{G}}=0$ V to the stated higher value of $V_{\mathrm{G}}$ are indicated by arrows.}
		\label{S3}
	\end{center}
\end{figure*}

The energy of the spin-orbit split valence band extrema and the linewidth at K$_{\mathrm{W}}$ for the spectra shown in Figs.  1(f)-(g) in the main text are determined by EDC fits as shown in Figs.  \ref{S3}(a)-(b).  In the case of vacuum exposed SL WS$_2$ we fit two Lorentzian peaks on a background described as a third order polynomial,  which provides excellent fits at both $V_{\mathrm{G}}=0$ V and at $V_{\mathrm{G}}=8$ V as shown in Fig.  \ref{S3}(a).  The extracted linewidth at the VBM is 0.17 eV at $V_{\mathrm{G}}=0$ V and 0.32 eV at $V_{\mathrm{G}}=8$ V.  The VBM exhibits a downwards shift of 0.61 eV.  

For graphene-encapsulated SL WS$_2$,  the fit is substantially more tricky because the intensity of the VB states is very low and situated on a highly non-linear background as seen in Fig.  \ref{S3}(b),  which stems from intensity spilling out from the graphene $\pi$-band nearby.  However, using Lorentzian peaks on a background described as a fifth order polynomial,  we are able to obtain excellent fits as shown in Fig.  \ref{S3}(b).  Note that at $V_{\mathrm{G}}=12$ V,  the intensity of the lower spin-split branch is not distinguishable from the graphene $\pi$-band intensity.  The extracted linewidth of the VBM is 0.16 eV at $V_{\mathrm{G}}=0$ V and 0.17 eV at $V_{\mathrm{G}}=12$ V, and the VBM exhibits a downwards shift of 0.80 eV. 

\begin{figure*} [h!]
	\begin{center}
		\includegraphics[width=0.5\textwidth]{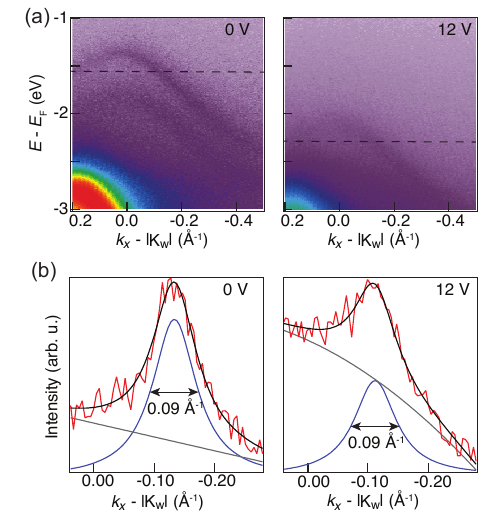}
		\caption{(a) ARPES spectra of valence bands around K$_{\mathrm{W}}$ at $V_{\mathrm{G}}=0$ V and $V_{\mathrm{G}}=12$ V. The high intensity in the lower lefter corner stems from the graphene $\pi$-band. (b) MDCs (red curves) with fits (black curves) to a single Lorentzian peak (blue curve) on a polynomial background (grey curve). The MDC is extracted 0.2 eV below the WS$_2$ VBM as shown via dashed horizontal lines in (a).}
		\label{S31}
	\end{center}
\end{figure*}

We extract a momentum distribution curve (MDC) 0.2 eV below the VBM of graphene-encapsulated WS$_2$ in order to check whether any momentum broadening occurs upon increasing $V_{\mathrm{G}}$. The MDC cut is shown via a dashed line in the ARPES spectra of the valence bands at $V_{\mathrm{G}}=0$ V and $V_{\mathrm{G}}=12$ V in Fig. \ref{S31}(a). Corresponding data with fits to a single Lorentzian peak on a polynomial background are shown in Fig. \ref{S31}(b). As $V_{\mathrm{G}}$ is increased and the valence band shifts to lower energies, the background in the MDC becomes increasingly more non-linear due to intensity spilling out from the graphene $\pi$-band (see lower left corner of the spectra in Fig. \ref{S31}(a)). At the extreme values of $V_{\mathrm{G}}$ of 0 V and 12 V, we find the Lorentzian full-width at half maximum remains fixed at 0.09 $\AA^{-1}$, which ascertains that there is no additional momentum broadening in the graphene-encapsulated WS$_2$ as we apply the gate voltage.

\newpage

\section{IV.  Analysis of conduction band dispersion}

The simple EDC analysis of CBM position presented in Fig. 2 and Fig.  3(c) of the main manuscript that lead to our band gap estimates is based on fitting a Lorentzian peak on a polynomial background multiplied by the Fermi-Dirac distribution.  The fitted EDC peak position is not necessarily a good estimate of the CBM position due to the relatively broad Fermi edge at room temperature combined with the finite EDC and MDC widths $\Delta E$ and $\Delta k$.  In the following we analyse how good an estimate of the CBM position this EDC peak position is using simulations of the two-dimensional $(E,k)$-dependent photoemission intensity \cite{Nechaev:2009}.  We simulate the intensity via the following expression:
\begin{eqnarray}
\mathcal{I}(E,k) = \left( \left[ \left(a+ bE\right) + \mathcal{I}_0\frac{\pi^{-1} \gamma}{(E-E_{\mathrm{c}}(k))^2 + \gamma^2} \right] (e^{(E-E_{\mathrm{F}})/k_BT}+1)^{-1}\right) \ast G(\Delta E) \ast  G(\Delta k), \nonumber
\label{eqn:1}
\end{eqnarray}
where the terms within the square brackets correspond to a background with a linear energy dependence and the spectral function of the WS$_2$ conduction band scaled by a constant intensity $\mathcal{I}_0$, which are multiplied by the Fermi Dirac distribution.  The linear background is assigned an offset $a$ and a slope $b$.  The spectral function is described by the constant $\gamma = 0.01$ eV and the conduction band dispersion $E_{\mathrm{c}}(k)$.  Furthermore,  $k_B$ is the Boltzmann constant and $T$ is the sample temperature, which is set to 300 K.  The expression within the large round brackets is convoluted by Gaussian functions $G$ with  EDC and MDC widths $\Delta E$ and $\Delta k$ that are set to the experimental values determined from EDC and MDC fits as described in Figs.  \ref{S3} and  \ref{S31}.  The conduction band dispersion is modelled as
\begin{eqnarray}
E_{\mathrm{c}}(k) = \frac{\hbar^2 k^2}{2m^{\ast}} - E_{\mathrm{min}},
\label{eqn:2}
\end{eqnarray}
where $m^{\ast}=0.31m_{\mathrm{e}}$ is the effective mass of the WS$_2$ conduction band expressed in units of the free electron mass $m_{\mathrm{e}}$ \cite{Shiang:2015}, and $E_{\mathrm{min}}$ is the position of the CBM.

\begin{figure*} [h!]
	\begin{center}
		\includegraphics[width=1\textwidth]{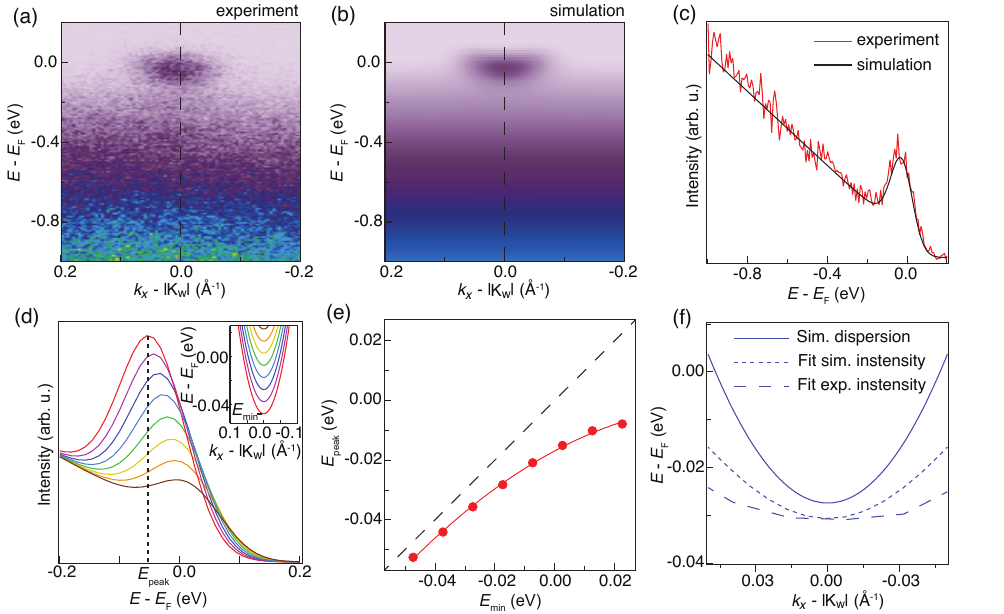}
		\caption{(a) ARPES data at $V_{\mathrm{G}} = 12$ V for graphene-encapsulated WS$_2$ as shown in Fig.  2(b) of the main manuscript.  (b) Simulated photoemission intensity for a parabolic band on a linear background adjusted to match the behavior of the intensity in (a).  (c) Comparison of EDC from experimental data and simulation at $k=0$ (see vertical dashed line in (a)-(b)).  (d) Simulated EDCs at $k=0$ for different shifts of the conduction band dispersion (see inset for corresponding dispersion curves that have the same color as the EDCs).  (e) Comparison of simulated EDC peak positions $E_{\mathrm{peak}}$ and CBM positions $E_{\mathrm{min}}$ (see (d) for definitions of $E_{\mathrm{peak}}$ and $E_{\mathrm{min}}$).  The dashed diagonal line illustrates an ideal 1:1 ratio between $E_{\mathrm{peak}}$ and $E_{\mathrm{min}}$. (f) Conduction band dispersion obtained from EDC fits of the experimental data (long-dashed curve) in (a) and from EDC fits of the simulated data (dotted curve) in (b) and raw dispersion used in the simulation (full curve).}
		\label{S3II}
	\end{center}
\end{figure*}

We adjust the parameters $a$,  $b$,  $\mathcal{I}_0$ and $E_{\mathrm{min}}$ in order to achieve an optimum fit of the EDC through the CBM at K$_{\mathrm{W}}$ between experiment and simulation.  A comparison of the measured ARPES spectrum and corresponding optimum simulation is presented via Figs.  \ref{S3II}(a)-(b) for graphene-encapsulated WS$_2$ at $V_{\mathrm{G}}=12$ V.  The resulting EDCs at K$_{\mathrm{W}}$ are shown in Fig.  \ref{S3II}(c),  demonstrating that an excellent fit can be achieved using our simple model of the intensity.  Next,  we proceed to vary $E_{\mathrm{min}}$ for our optimum parameters and thus simulate the effect on the EDC at  K$_{\mathrm{W}}$ of shifting the CBM,  as shown in Fig.  \ref{S3II}(d).  From the EDCs,  we extract the EDC peak position $E_{\mathrm{peak}}$,  which is compared with $E_{\mathrm{min}}$ for our parameters in Fig.  \ref{S3II}(e).  If $E_{\mathrm{peak}}$ and $E_{\mathrm{min}}$ were identical,  the markers would follow a linear proportional relationship with a slope of 1 (see dashed line in Fig.  \ref{S3II}(e)).  We observe that as $E_{\mathrm{min}}$ shifts to lower energies the value of $E_{\mathrm{peak}}$ approaches $E_{\mathrm{min}}$,  whereas the deviation between the two values reaches 20 meV around $E_{\mathrm{F}}$ and increases towards higher energies.  However,  for our CBM estimates in Fig.  3(c) of the main paper any such discrepancy between our fitted EDC peak position and the actual position of the CBM would be well within our error bar of the band gap of $\pm 0.03$ eV.

\begin{figure*} [h!]
	\begin{center}
		\includegraphics[width=1\textwidth]{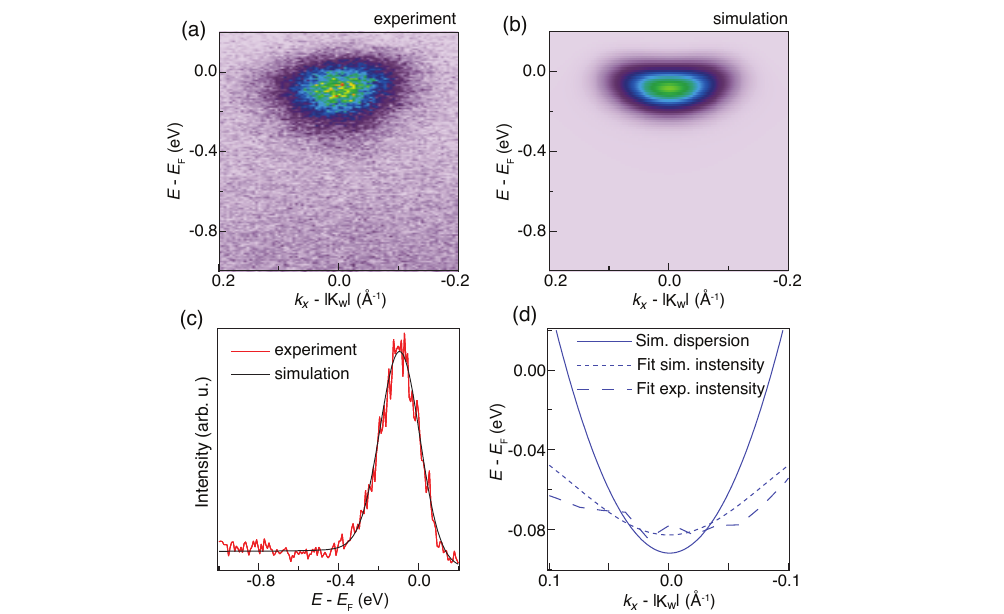}
		\caption{(a) ARPES data at $V_{\mathrm{G}} = 8$ V for vacuum exposed WS$_2$ as shown in Fig.  2(a) of the main manuscript.  (b) Simulated photoemission intensity adjusted to the measured intensity in (a). (c) Comparison of measured and simulated EDCs at $k=0$.  (d) Simulated dispersion (full curve) and dispersion curves obtained from EDC fits of experiment (long-dashed curve) and simulation (dotted curve).}
		\label{S3III}
	\end{center}
\end{figure*}

We comment on the possibility to extract the conduction band dispersion and effective mass from the ARPES intensity.  A common approach involves EDC fits of the ARPES intensity of the populated conduction band states at $k$-values around $E_{\mathrm{F}}$ and using the resulting EDC peak position at each $k$-value to re-create the dispersion and extract $m^{\ast}$.  However,  due to the deviation between the raw dispersion and the energy of the peak intensity described above,  this leads to an apparent effective mass $m^{\ast}_{\mathrm{apparent}}$ with a significantly different value than the real $m^{\ast}$ of the underlying dispersion.  This is demonstrated in Fig.  \ref{S3II}(f) by comparing the simulated dispersion (full curve) where $m^{\ast}=0.31m_{\mathrm{e}}$ with EDC fits of the simulated dispersion (dotted curve),  which exhibits an apparent flattening with $m^{\ast}_{\mathrm{apparent}}=0.61m_{\mathrm{e}}$.  Interestingly,  EDC fits of the experimental data reveal an even flatter dispersion (long-dashed curve) with $m^{\ast}_{\mathrm{apparent}}=1.50m_{\mathrm{e}}$,  which could indicate a renormalization effect due to electron-plasmon coupling, as discussed in connection with our $G_0W_0$ analysis \cite{ulstrup2024observation}.

Figure \ref{S3III} presents a similar analysis of the conduction band of vacuum exposed WS$_2$ at $V_{\mathrm{G}}=8$ V.  For this data, we also obtain an excellent fit,  as demonstrated via Figs. \ref{S3III}(a)-(c), by only adjusting the parameters relating to the magnitude of the intensity and $E_{\mathrm{min}}$.  Here,  the CBM is shifted significantly below $E_{\mathrm{F}}$ such that $E_{\mathrm{peak}}$ provides a good description of $E_{\mathrm{min}}$.  This is seen via the close proximity of the minima of the conduction band dispersion curves from simulation and EDC fits in  Fig.  \ref{S3III}(d).  Note that the CBM is slightly underestimated via the EDCs,  by less than 10 meV,  due to the energy broadening stemming from the finite photocurrent in this section of the device.  In addition,  this broadening significantly flattens the dispersion which leads to $m^{\ast}_{\mathrm{apparent}}$ exceeding $1.00m_{\mathrm{e}}$, as determined via EDC fits of both simulated and measured data.

\newpage

\section{V. Gate voltage dependence of band gaps}

\begin{figure*} [h!]
	\begin{center}
		\includegraphics[width=1\textwidth]{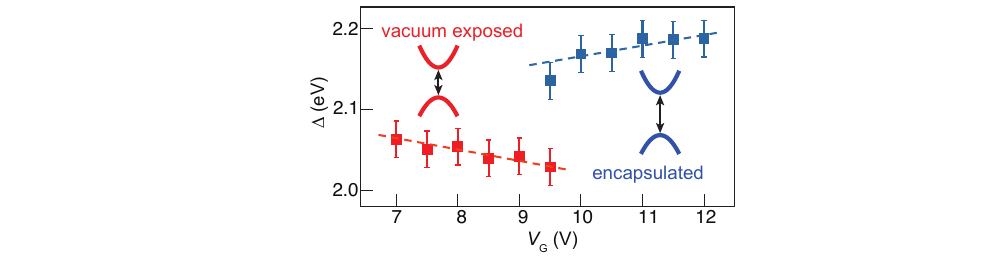}
		\caption{$V_{\mathrm{G}}$-dependence of band gap extracted from the VBM and CBM positions in Fig. 3(c) of the main manuscript.  Dashed lines represent linear fits that serve as guides to the eye.  The insets illustrate the band gap renormalization between vacuum exposed and graphene-encapsulated SL WS$_2$.}
		\label{S4}
	\end{center}
\end{figure*}

Band gap values $\mathrm{\Delta}$ extracted as a function of $V_{\mathrm{G}}$ from our $V_{\mathrm{G}}$-sweeps in Fig. 3(c) of the main manuscript are shown in Fig. \ref{S4} for the two section of the WS$_2$. The gap values stated in the main text represent the average and the error represents the standard deviation of $\mathrm{\Delta}$ at all $V_{\mathrm{G}}$-points in a given section. Note that in the vacuum exposed part we see that the gap appears to tend to a lower value as we increase $V_{\mathrm{G}}$. This is sensible as we may slightly increase the doping and thus the internal screening, which will further decrease the gap. For graphene-encapsulated WS$_2$ we see that the gap, surprisingly, appears to increase as $V_{\mathrm{G}}$ increases. This behavior may arise from the underestimation of the energy of the CBM via the EDC analysis at low doping, as discussed in connection with Fig.  \ref{S3II}(e).  It is also possible that the complex doping dependence of the photoemission intensity composed of quasiparticle and plasmon satellite peaks could impact the extracted gap. Given our energy resolution and the magnitude of the effects, however, we can only remain speculative about such possible trends, which warrant further detailed experimental studies.

\newpage

\section{VI. Band positions at the vacuum-graphene interface}

\begin{figure*} [h!]
	\begin{center}
		\includegraphics[width=1\textwidth]{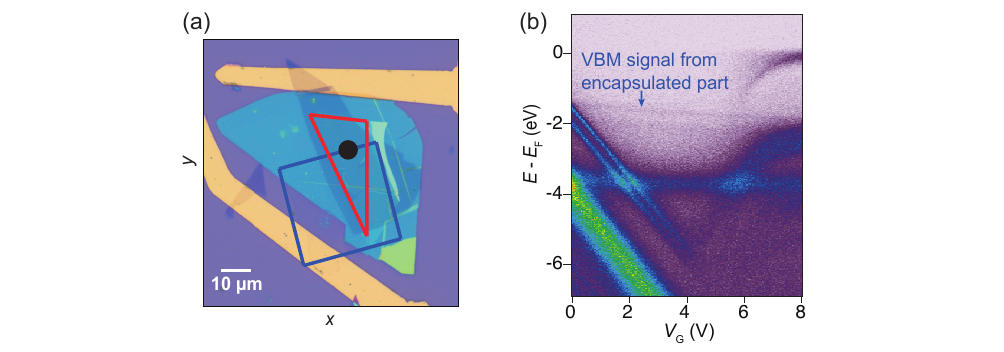}
		\caption{(a) Optical micrograph of our device as shown in Fig. 1(b) of the main manuscript. The black circle represents the light spot position for a microARPES measurement at the interface between the vacuum exposed and graphene-encapsulated sections of SL WS$_2$. (b) EDC at K$_{\mathrm{W}}$ as a function of $V_{\mathrm{G}}$ from 0 to 8 V measured at the interface. The arrow demarcates faint valence band signal stemming from the graphene-encapsulated side of WS$_2$.}
		\label{S5}
	\end{center}
\end{figure*}

In the main text we discuss VBM and CBM positions in vacuum exposed and graphene-encapsulated sections of our SL WS$_2$ device. An intriguing question that emerges during such analysis is whether the band positions exhibit any spatial dependence - i.e. lateral band bending effects - that we can resolve with microARPES across the device. Since our light spot is placed deep into the two separate sections of the device in the analysis in Fig. 3 of the main manuscript, we can investigate for any spatial dependence by placing the light spot in the middle of the interface between the two sections, as shown via the black circle in the optical micrograph in Fig. \ref{S5}(a). Band positions are measured in this location via an EDC at K$_{\mathrm{W}}$ as a function of $V_{\mathrm{G}}$ from 0 to 8 V, as shown in Fig. \ref{S5}(b). We observe a set of valence bands that rigidly shift downwards from 0 to 5 V, following the gate potential, which then shift back up above $V_{\mathrm{G}} = 5$ V as the CBM becomes populated (and shifts up together with the valence bands). In addition, we find a faint intensity from another set of valence bands, which shift monotonously down in energy as $V_{\mathrm{G}}$ is increased. The features represent a juxtaposition of the photoemission intensities we measure in the separate sections of the device shown in Figs. 3(a)-(b) of the main manuscript. As we observe the same band positions and $V_{\mathrm{G}}$-dependence at the interface, we can conclude that any lateral band bending is occuring on a substantially smaller length scale than the 4 $\mu$m size of our light spot.

\newpage

\section{VII.  Model Ab-Initio $G_0W_0$ Calculations}
We perform our $G_0W_0$ calculations based on an ab initio derived minimal lattice many-body model for SL WS$_{2}$. This model is formulated in a minimal basis of the three dominant d orbitals (around the undoped band gap) and includes kinetic hopping terms as well as non-local Coulomb interaction matrix elements. The corresponding Hamiltonian is
\begin{equation}
    H = \sum_{\mathbf{k},a,b,\sigma} \varepsilon_{ab\sigma}(\mathbf{k}) c^\dagger_{a\sigma}(\mathbf{k}) c^{\phantom{\dagger}}_{b\sigma}(\mathbf{k})
    + \sum_{\mathbf{k},\mathbf{k}',\mathbf{q},a,b,c,d,\sigma \sigma'} U_{abcd}(\mathbf{q}) c^\dagger_{a\sigma}(\mathbf{k}+\mathbf{q}) c^\dagger_{b\sigma'}(\mathbf{k}'-\mathbf{q}) c^{\phantom{\dagger}}_{c\sigma'}(\mathbf{k}') c^{\phantom{\dagger}}_{d\sigma}(\mathbf{k}),
\end{equation}
where $\mathbf{k}$, $\mathbf{k}'$ and $\mathbf{q}$ are crystal (transfer) momenta, $a$, $b$, $c$ and $d$ are orbital indices in the minimal three-orbital basis, and $\sigma$ and $\sigma'$ are spin indices. $\varepsilon_{ab\sigma}(\mathbf{k})$ are the Fourier transformed hopping elements and $U_{abcd}(\mathbf{q})$ are the Coulomb matrix elements.
Using this model allows us to introduce the environmental screening (using our Wannier function continuum electrostatics approach \cite{rosner2015wannier}) as well as to adjust the Fermi level to mimic electron doping.  

For the derivation of the minimal lattice many-body model we use a combination of density functional theory (DFT) calculations, Wannier function basis sets, and constrained random phase approximation (cRPA) calculations\cite{aryasetiawan2004frequency-dependent}. We start from a DFT calculation for freestanding and undoped monolayer WS$_{2}$ using the Vienna Ab Initio Simulation Package (VASP)\cite{kresse1996efficient,kresse1999ultrasoft} using a projector augmented planewave (PAW) basis set\cite{blochl1994projector} and a PBE exchange correlation functional\cite{perdew1996generalized}. The lattice constant of WS$_2$ is set to 3.186~\AA. For simplicity we neglect spin-orbit coupling effects, as these would only affect our results quantitatively, but not the qualitative trends we are interested in. Afterwards we project the Kohn-Sham bands to Mo $d_{z^2}, d_{xy}, d_{x^2-y^2}$ orbitals using Wannier 90. From this we get the minimal tight-binding model, which captures all details of the full ab initio band structure.  With these Wannier functions, we perform cRPA calculations to obtain partially screened Coulomb interaction matrix elements, which take the screening of all other bands (except those with predominant $d_{z^2}, d_{xy}, d_{x^2-y^2}$ character) into account. The calculations are done in momentum space using $16 \times 16  \times 1$ $k$/$q$ grids and 192 bands. The resulting Coulomb interaction data points are used in a last step to fit an analytical model for the density-density Coulomb matrix elements. To this end, we diagonalize the cRPA density-density Coulomb matrix elements, resulting in 3 eigenvalues. Similar to Ref.~\cite{rosner2015wannier}, the leading eigenvalue $U_{1}(\mathbf{q})$ is fitted to the following function
\begin{equation}
    U_{1}(\mathbf{q}) = \frac{4 e^2}{\tilde{A} \varepsilon_{\text{back}}(\mathbf{q})} \frac{\text{arctan}\left( \frac{\pi}{q d} \right)}{q},
    \label{eqSI:CoulombFit}
\end{equation}
with $d = 6.22$\,$\AA$ the effective material thickness, $e$ the electron charge and $\tilde{A} = A/3$ where is $A$ the unit-cell area.
The singularity of this function at $\mathbf{q}=0$ should be canceled by the positive charge background. To account for this, we fit the value $U_1(\mathbf{q}=0)$ such that our $G_0W_0$ results in the minimal model are in reasonable agreement with other ab-initio $G_0W_0$ results for the gap of an undoped, freestanding monolayer WS$_2$\cite{haastrup2018computational,gjerding2021recent}. 
At non-zero $\mathbf{q}$, screening from all bands outside the minimal model, as well as environmental screening, is captured by the background dielectric function\cite{rosner2015wannier}
\begin{equation}
    \varepsilon_{\text{back}}(\mathbf{q}) = \varepsilon_{\text{int}}(\mathbf{q}) \frac{1 - \beta(\mathbf{q}) e^{-q d}}{1 + \beta(\mathbf{q}) e^{-q d}},
\end{equation}
with $\beta(\mathbf{q}) = (\varepsilon_{\text{int}}(\mathbf{q}) - \varepsilon_{\text{env}}) / (\varepsilon_{\text{int}}(\mathbf{q}) + \varepsilon_{\text{env}})$ and the internal dielectric function $\varepsilon_{\text{int}}(\mathbf{q})$ described by\cite{steinhoff_exciton_2017,florian2018dielectric}
\begin{equation}
    \varepsilon_{\text{int}}(\mathbf{q}) = \frac{A + q^2}{A \text{sin}(q C) / (q B C) + q^2} + E.
\end{equation}
Here $A = 0.42$\,$\AA^{-2}$, $B = 2.10$, $C = 5.58$\,$\AA$, $E = 2.95$ and $\varepsilon_{\text{env}} = 1.59$ are fitting parameters. After the fitting procedure, the dielectric constant $\varepsilon_{\text{env}}$ is treated as a tuning parameter to describe different dielectric environments.
Since the sub-leading eigenvalues of the cRPA density-density Coulomb matrix elements have a relatively weak momentum dependence\cite{rosner2015wannier,steinhoff_exciton_2017}, they are approximated by their average value $U_{2,3}(\mathbf{q}) \approx U_{2,3} = \sum_{\mathbf{q}} U_{2,3}(\mathbf{q}) / N_q$, with $N_q$ the number of q-grid points.

Equipped with an analytical Coulomb model, the subsequent real-frequency $G_0W_0$ calculations are performed on the minimal tight-binding model, while retaining the full frequency and momentum dependence of the self-energy in order to describe the plasmon satellite features. These calculations are performed within the TRIQS \cite{parcollet2015triqs} and TPRF \cite{strand2023tprf} codebases, using 1200 frequency points in an energy window of $\pm$ 9 eV around the center of the (undoped) gap, in combination with $64 \times 64$ $k$/$q$ grids. The doping is adjusted by changing the Fermi level, which affects the bare polarization and thereby the screening of the Coulomb interaction. 
% The amount of doping is denoted by $\delta_{\mathrm{e}}$, which is defined as the amount of additional charge carriers per unit-cell area.

We note that the discrepancies between theoretical and experimental changes of the gap size between vacuum exposed and graphene-encapsulated WS$_2$ could in part be due to the underlying simplified WS$_2$ many-body model, the approximate handling of the environmental screening [here described by a dielectric constant $\varepsilon_{\mathrm{env}}$, which is in reality a complex function $\varepsilon_{\mathrm{env}}(\mathbf{q},\omega)$], and/or the missing diagrams in the $G_0W_0$ approximation. Regarding the level of theory, it is, for example, known that vertex corrections within the so-called GW+cumulant scheme or via excitonic screening channels quantitatively affect $\omega_{\mathrm{pl}}(\mathrm{K_W})$ and quasiparticle gaps\cite{aryasetiawan1996multiple,kas2014cumulant,caruso2015band,steinhoff_exciton_2017}. Thus, we cannot exclude their relevance for the gated and partially graphene covered WS$_2$ heterostructure here.

\end{document}